\documentclass[%
 reprint,
nofootinbib,
 amsmath,amssymb,
 aps,
floatfix,
]{revtex4-1}

\usepackage{graphicx}% Include figure files
\usepackage{dcolumn}% Align table columns on decimal point
\usepackage{bm}% bold math
\usepackage{appendix}
\usepackage{hyperref}% add hypertext capabilities
\usepackage{adjustbox}
\usepackage{xcolor}

\newcommand{\ba}{\begin{eqnarray}}
\newcommand{\ea}{\end{eqnarray}}

\newcommand{\be}{\begin{equation}}
\newcommand{\ee}{\end{equation}}

\begin{document}

%\preprint{APS/123-QED}
%\preprint{LTH-1291}

\title{$Z'$s and sterile neutrinos from heterotic string models: \\ \medskip
exploring $Z'$ mass exclusion limits }% Force line breaks with \\

\author{Alon E. Faraggi$^{(a,b,1)}$, Marco Guzzi$^{(c,2)}$}
\email{$^{1}$Alon.Faraggi@liverpool.ac.uk, $^{2}$mguzzi@kennesaw.edu}

\affiliation{$^{(a)}$Department of Mathematical Sciences
University of Liverpool, Liverpool L69 7ZL, UK. \\
$^{(b)}$Department of Particle Physics and Astrophysics, Weizmann Institute,
Rehovot 76100, Israel.\\
$^{(c)}$Department of Physics, Kennesaw State University, 370 Paulding Ave.,  30144 Kennesaw, GA, U.S.A.}

\date{\today}% It is always \today, today,

\begin{abstract}
\begin{description}
\item[Abstract]
We investigate the impact of sterile neutrinos on the decay rate
of extra $Z'$s with mass in the TeV range in heterotic string derived models.
We explore the impact of sterile neutrinos on the current $Z'$ mass exclusion limits at the LHC, and how these bounds change when the parameter space of this specific class of models is modified.

%\item[Pacs numbers] 12.38.-t,12.38.Bx,12.38.Aw
\end{description}
\end{abstract}

\maketitle

\tableofcontents

\section{\label{sec:Introduction} Introduction}

The Standard Model of particle physics provides an effective quantum field theory parameterisation of
all observational data to date. Furthermore, the logarithmic evolution of the Standard Model
parameters suggests that it may provide such viable parameterisation up to the Grand Unified Theory
scale, or the Planck scale. Nevertheless, it is in general anticipated that the Standard Model is
augmented by additional particles and symmetries due to its numerous shortcomings, {\it e.g.}: it does
not provide viable sources for dark matter and dark energy; it is composed of several ad hoc sectors
and requires 19 continuous, and numerous discrete, parameters to account for the experimental data; it
does not account for the gravitational interactions, which is fundamentally incompatible with the
quantum field theory framework. Chief among them is the fact that neutrinos are massless in the renormalisable Standard Model, which contradicts the experimental observations.
It is clear that the neutrino sector of the Standard Model and its extensions provide the most fertile
ground for experiments in the near future.
Perhaps most intriguing in this context is the possible existence of light sterile neutrinos.
In turn, it was argued that sterile neutrinos can exist at low scales provided that they are charged under an extra $U(1)$ gauge symmetry, which remains
unbroken down to low scales, and under which the sterile neutrinos are chiral. Mass terms for the
sterile neutrinos can only be generated by the Vacuum Expectation Value (VEV) that breaks the extra
$U(1)$ symmetry. Otherwise, absence of global symmetries in quantum gravity, in general, leads to the
expectation that Planck scale mass terms will be generated for particles whose mass scales are not
protected by a chiral symmetry. We may allow for their mass terms to be suppressed by several orders
of magnitude relative to the Planck scale \cite{Faraggi:1993rc}, but suppression of their masses to
the GeV scale, or below, requires high degree of fine tuning, or a remnant local discrete symmetry
\cite{Faraggi:1996yu}. This general expectation is indeed borne out in explicit quasi--realistic
string constructions \cite{Faraggi:1993zh, Coriano:2003ui}. Light sterile neutrinos are therefore
naturally associated with an extra $U(1)$ gauge symmetry that may be within reach of the LHC
\cite{Faraggi:2018pit}.
On the other hand, the existence of light sterile neutrinos may have profound impact on the collider
signatures of the extra vector boson, as it may substantially affect its branching ratios, compared
to extra $Z^\prime$ models without sterile neutrinos. In this paper we examine the potential impact of
the sterile neutrinos on the experimental signatures of the extra string derived $Z^\prime$ model. We
remark that construction of string models with extra $Z^\prime$ gauge symmetries, that may remain
unbroken down to low scales, has proven to be an arduous task. The reason is that the extra $U(1)$
symmetries that are often discussed in the context of GUT extensions of the Standard Model are
anomalous in explicit string models and therefore cannot remain unbroken down to low scales.
In this paper we examine the potential impact of light sterile neutrinos on the signature of extra $Z^\prime$ vector boson in string derived models.
Related literature can be found in refs.~\cite{Komachenko:1989qn,Rizzo:1998ut,Arun:2022ecj,Das:2017deo,Das:2017flq,Das:2019fee,Chiang:2019ajm,Das:2022rbl,Das:2021esm,Anastasopoulos:2008jt,Anastasopoulos:2022sji}.

In section~\ref{zprecap} we give a brief overview of the
construction of string models with extra $Z^\prime$ gauge symmetry.
In section~\ref{Neutral-sector} we analyze the neutral gauge boson sector of the model and
in section~\ref{decay-rates} we study the decay rates and analyze the branching ratios of the extra $Z^\prime$. In section\ref{sterile-neutrinos} we investigate the sterile neutrino sector and calculate the branching ratios of the $Z'$ into sterile neutrinos and compare with the corresponding rates in their absence. Finally, in section~\ref{collider-signatures} we study phenomenological implications and signatures at hadron colliders. Our concluding remarks are in section~\ref{conclusions}.

\section{Extra $Z^\prime$ in string derived models}\label{zprecap}
In this section we discuss the structure of the string derived $Z^\prime$ model
of ref.~\cite{Faraggi:2014ica}. The model was constructed in the free fermionic
formulation \cite{Antoniadis:1986rn,Kawai:1986ah,Antoniadis:1987wp}.
%The details of the constructions will not be repeated here, and
Only the
most salient features relevant for our discussion are highlighted here.
In the free fermionic formulation in four dimensions
all the worldsheet degrees of freedom required to cancel the conformal anomaly
are represented in terms of free fermions
on the string worldsheet. In a convenient notation the 64 worldsheet
fermions in the lightcone gauge are denoted as:
\leftline{${\underline{{\hbox{Left-Movers}}}}$:~~$\psi^\mu,~~{ \chi_i},
~~{ y_i,~~\omega_i}~~~~(\mu=1,2,~i=1,\cdots,6)$}
\vspace{4mm}
\leftline{${\underline{{\hbox{Right-Movers}}}}$}
$${\bar\phi}_{A=1,\cdots,44}=
\begin{cases}
~~{ {\bar y}_i~,~ {\bar\omega}_i} & i=1,{\cdots},6\cr
  & \cr
~~{ {\bar\eta}_i} & i=1,2,3~~\cr
~~{ {\bar\psi}_{1,\cdots,5}} & \cr
~~{{\bar\phi}_{1,\cdots,8}}  &
\end{cases}
$$
where the $\{y,\omega\vert{\bar y},{\bar\omega}\}^{1,\cdots,6}$
correspond to the
six compactified dimensions of the internal space;
${\bar\psi}^{1,\cdots,5}$ produce the $SO(10)$ GUT symmetry;
${\bar\phi}^{1,\cdots,8}$ produce the
hidden sector gauge group; and ${\bar\eta}^{1,2,3}$
produce three $U(1)$ gauge symmetries.
Models in the free fermionic formulation are specified in terms of boundary
condition basis vectors, which denote the transformation properties
of the fermions around the noncontractible loops of the worldsheet torus,
and the Generalised GSO projection coefficients of the one loop
partition function
\cite{Antoniadis:1986rn, Kawai:1986ah, Antoniadis:1987wp}.
The free fermion models correspond to toroidal
$Z_2\times Z_2$ orbifolds with discrete Wilson lines
\cite{Athanasopoulos:2016aws}.

Interest in string inspired $Z^\prime$ models arose from the
discovery that string inspired effective field theory models
give rise to $E_6$ GUT like models \cite{Candelas:1985en}.
Extra $U(1)$ symmetries
in these string inspired models therefore possess an $E_6$ embedding
and have generated multitude of papers since the mid--eighties
(for a recent review see {\it e.g.} \cite{King:2020ldn}).
The construction of string derived models that admit an unbroken
extra $U(1)$ symmetry down to low scales proves, however,
to be very difficult.
The symmetry breaking pattern in the string
models $E_6\rightarrow SO(10)\times U(1)_A$ entails that $U(1)_A$
is anomalous and cannot be part of a low scale unbroken $U(1)_{Z^\prime}$
\cite{Cleaver:1997rk}.
String derived constructions with low scale
$U(1)_{Z^\prime}\notin E_6$ were analysed in
\cite{Pati:1996fn, Faraggi:2000cm, Coriano:2007ba, Faraggi:2011xu},
but agreement with the measured values of $\sin^2(\theta)_W(M_Z)$ and
$\alpha_s(M_Z)$ favours $Z^\prime$ models with $E_6$ embedding
\cite{Faraggi:2013nia}.
We note that the anomaly free $U(1)$ combination of
$U(1)_{B-L}$ and $U(1)_{T_{3_R}}\in SO(10)$ may in principle
remain unbroken down to low scales \cite{Faraggi:1990ita}.
However, ensuring that the left--handed neutrino masses are adequately
suppressed is facilitated if this $U(1)$ symmetry is broken at a high
scale \cite{Faraggi:1990it}. Constructing string models that allow for
a extra $U(1)\in E_6$ symmetry to remain unbroken down to low scales
necessitates the construction of string models in which
$U(1)_A$ is rendered anomaly free. One route to achieving this outcome
is to enhance $U(1)_A$ to a non-Abelian gauge symmetry {\'a} la
ref. \cite{Bernard:2012vf}.
An alternative is the construction of ref. \cite{Faraggi:2014ica},
which utilises the spinor--vector duality that was observed in
$Z_2\times Z_2$ orbifolds
\cite{Faraggi:2006pk,Angelantonj:2010zj, Faraggi:2011aw}.
The duality is under the
exchange of the total number of $(16+\overline{16})$ representations
of $SO(10)$ with the total number of $10$ representations, and
is easy to understand if we consider the extension of
$SO(10)\times U(1)_A$ to $E_6$.
The chiral and anti--chiral representations
of $E_6$ decompose under $SO(10)\times U(1)$ as $27=16+10+1$ and
$\overline{27}= \overline{16}+10+1$.
In this case the $\#_1$ of $(16+\overline{16})$ and $\#_2$ of $10$
representations are equal.
The $E_6$ symmetry point in the moduli space
corresponds to a self--dual point under the exchange
of the total number of $SO(10)$ spinorial plus anti--spinorial, with the
total number of vectorial, representations.
Breaking the $E_6$ symmetry to $SO(10)\times U(1)_A$ results in the projection
of some of the spinorial and vectorial representations, which renders $U(1)_A$
anomalous. However, there may exist vacua with equal numbers of
$(16+\overline{16})$ spinorial, and $10$ vectorial, representations,
and traceless $U(1)_A$, without enhancement of the $SO(10)\times U(1)_A$
symmetry to $E_6$ \cite{Faraggi:2014ica}.
In such models the chiral spectrum still forms
complete $E_6$ representations, but the gauge symmetry is not
enhanced to $E_6$. In this cases
$U(1)_A$ may be anomaly free and remain unbroken down to low
scales.

A classification method that provides a fishing tool to extract models
with specified physical properties was developed by using the free fermionic
model building rules
\cite{Faraggi:2004rq, Assel:2010wj,
  Faraggi:2014hqa,Faraggi:2017cnh,
  Faraggi:2018hqx, Faraggi:2019qoq,
  Faraggi:2020wej,Faraggi:2020wld}.
In ref. \cite{Faraggi:2014ica}, using the free fermionic fishing algorithm, such
a spinor--vector self dual model was extracted with subsequent breaking
of the $SO(10)$ symmetry to $SO(6)\times SO(4)$, which preserves the
spinor--vector self--duality. This model is a string derived model
in which an extra $U(1)$ with $E_6$ embedding may remain unbroken down to
low scales. The full massless spectrum of the string derived model is given
in ref. \cite{Faraggi:2014ica}.
The observable and hidden gauge groups at the
string scale are produced by untwisted sector states and are given by:
\begin{eqnarray}
{\rm observable} ~: &~~SO(6)\times SO(4) \times
U(1)_1 \times U(1)_2\times U(1)_3 \nonumber\\
{\rm hidden}     ~: &SO(4)^2\times SO(8)~~~~~~~~~~~~~~~~~~~~~~~~~~~~~~~\nonumber
\end{eqnarray}
The massless string spectrum contains the
fields required to break the GUT symmetry to the Standard Model.
There are two anomalous
$U(1)$s in the string model with
\begin{equation}
{\rm Tr}U(1)_1= 36 ~~~~~~~{\rm and}~~~~~~~{\rm Tr}U(1)_3= -36.
\label{u1u3}
\end{equation}
The $E_6$ combination, given by
\begin{equation}
U(1)_\zeta ~=~ U(1)_1+U(1)_2+U(1)_3~,
\label{uzeta}
\end{equation}
is anomaly free
and can be a component of an extra $Z^\prime$ below the string scale.
The observable $SO(6)\times SO(4)$ gauge symmetry in the model is
broken by the VEVs of the heavy Higgs fields ${\cal H}$ and
$\overline{\cal H}$ . The decomposition of these fields in terms of the
Standard Model gauge group factors is given by:
\begin{align}
\overline{\cal H}({\bf\bar4},{\bf1},{\bf2})& \rightarrow u^c_H\left({\bf\bar3},
{\bf1},\frac 23\right)+d^c_H\left({\bf\bar 3},{\bf1},-\frac 13\right)+\nonumber\\
         &   ~~~~~~~~~~~~~~~~~~  {\overline {\cal N}}\left({\bf1},{\bf1},0\right)+
                             e^c_H\left({\bf1},{\bf1},-1\right)
                             \nonumber \\
{\cal H}\left({\bf4},{\bf1},{\bf2}\right) &
\rightarrow  u_H\left({\bf3},{\bf1},-\frac 23\right)+
d_H\left({\bf3},{\bf1},\frac 13\right)+\nonumber\\
         &  ~~~~~~~~~~~~~~~~~~   {\cal N}\left({\bf1},{\bf1},0\right)+
e_H\left({\bf1},{\bf1},1\right)\nonumber
\end{align}
The VEVs along the ${\cal N}$ and $\overline{\cal N}$ directions leave the
unbroken combination
\begin{equation}
U(1)_{{Z}^\prime} ~=~
\frac{1}{5} (U(1)_C - U(1)_L) - U(1)_\zeta
~\notin~ SO(10),
\label{uzpwuzeta}
\end{equation}
that may remain unbroken below the string scale provided that $U(1)_\zeta$ is
anomaly free. We note that this symmetry breaking pattern is enforced
in the string model due
to a doublet--triplet missing partner mechanism \cite{Antoniadis:1988cm}
that gives heavy mass
to coloured scalar states that arise in the untwisted sector of the string
model \cite{Faraggi:2014ica}. Thus, the combination given in eq.
(\ref{uzpwuzeta}) is the extra $U(1)$ combination that can arise
in the string derived model with $E_6$ embedding of the charges.
Anomaly cancellation of the $U(1)_{Z^\prime}$ charges requires
the existence of the vector--like leptons
$\{H_{\textrm{vl}}^i, {\bar H}_{\textrm{vl}}^i\}$, and quarks $\{D^i, \overline{D}^i\}$, that arise
from the vectorial $10$ representation of $SO(10)$, as well as the $SO(10)$
singlets $S^i$ in the $27$ of $E_6$.
The supermultiplet\footnote{Superfields are indicated with a hat symbol.}
spectrum below the Pati--Salam breaking
scale is displayed schematically in Table~\ref{table27rot}.
The three right--handed
neutrino $N_L^i$ states become massive at the $SU(2)_R$ breaking scale,
which generates the seesaw mechanism \cite{Faraggi:2018pit}.
The spectrum below the $SU(2)_R$ breaking scale is assumed to be
supersymmetric, and we therefore include in the spectrum
an additional pair of vector--like electroweak Higgs doublets,
that facilitate gauge coupling unification.
This is justified in the string inspired models due to the string
doublet--triplet splitting mechanisn \cite{Faraggi:1994cv, Faraggi:2001ry}.
The states $\phi$ and ${\bar\phi}$ are
exotic Wilsonian states \cite{Faraggi:2014ica, DelleRose:2017vvz}.
Additionally, the existence of light states $\zeta_i$,
that are neutral under the
$SU(3)_C\times SU(2)_L\times U(1)_Y\times U(1)_{Z^\prime}$ low
scale gauge group, is allowed.
The $U(1)_{Z^\prime}$
gauge symmetry can be broken at low scales by the VEV of the $SO(10)$ singlets
$S_i$ and/or ${\phi_{1,2}}$.

\begin{table}[!ht]
\noindent
{\small
\begin{center}
{
%\tabulinesep=1.2mm
\begin{tabular}{|l|cc|c|c|c|}
\hline
Field &$\hphantom{\times}SU(3)_C$&$\times SU(2)_L $
&${U(1)}_{Y}$&${U(1)}_{Z^\prime}$  \\
%\hline
\hline
$\hat{Q}_L^i$&    $3$       &  $2$ &  $+\frac{1}{6}$   & $-\frac{2}{5}$   ~~  \\
$\hat{u}_L^i$&    ${\bar3}$ &  $1$ &  $-\frac{2}{3}$   & $-\frac{2}{5}$   ~~  \\
$\hat{d}_L^i$&    ${\bar3}$ &  $1$ &  $+\frac{1}{3}$   & $-\frac{4}{5}$  ~~  \\
$\hat{e}_L^i$&    $1$       &  $1$ &  $+1          $   & $-\frac{2}{5}$  ~~  \\
$\hat{L}_L^i$&    $1$       &  $2$ &  $-\frac{1}{2}$   & $-\frac{4}{5}$  ~~  \\
\hline
$\hat{D}^i$       & $3$     & $1$ & $-\frac{1}{3}$     & $+\frac{4}{5}$  ~~    \\
$\hat{{\bar D}}^i$& ${\bar3}$ & $1$ &  $+\frac{1}{3}$  &   $+\frac{6}{5}$  ~~    \\
$\hat{H}_{\textrm{vl}}^i$       & $1$       & $2$ &  $-\frac{1}{2}$   &  $+\frac{6}{5}$ ~~    \\
$\hat{{\bar H}}_{\textrm{vl}}^i$& $1$       & $2$ &  $+\frac{1}{2}$   &   $+\frac{4}{5}$   ~~  \\
\hline
$\hat{S}^i$       & $1$       & $1$ &  ~~$0$  &  $-2$       ~~   \\
\hline
$\hat{H}_1$         & $1$       & $2$ &  $-\frac{1}{2}$  &  $-\frac{4}{5}$  ~~    \\
$\hat{H}_2$  & $1$       & $2$ &  $+\frac{1}{2}$  &  $+\frac{4}{5}$  ~~    \\
\hline
$\hat{\phi}$       & $1$       & $1$ &  ~~$0$         & $-1$     ~~   \\
$\hat{\bar\phi}$       & $1$       & $1$ &  ~~$0$     & $+1$     ~~   \\
\hline
\hline
$\hat{\zeta}^i$       & $1$       & $1$ &  ~~$0$  &  ~~$0$       ~~   \\
\hline
\end{tabular}}
\end{center}
}
\caption{\label{table27rot}
\it
Supermultiplet spectrum and
$SU(3)_C\times SU(2)_L\times U(1)_{Y}\times U(1)_{{Z}^\prime}$
quantum numbers, with $i=1,2,3$ for the three light
generations. The charges are displayed in the
normalisation used in free fermionic
heterotic--string models. }
\end{table}

\section{Neutral gauge bosons sector}
\label{Neutral-sector}
The scalar Lagrangian describing the mass contributions of the gauge bosons is given by
\ba
{\cal L}_{\textsc{GM}}=\vert{\mathcal D}_{\mu}H_1\vert^{2}
+\vert{\mathcal D}_{\mu}H_2\vert^{2}
+\sum_{i}\vert{\mathcal D}_{\mu}S_i\vert^{2}
\label{scalar-sec}
\ea
where the scalar components of the supermultiplets are $S_i$, and $H_1$ and $H_2$ which are the singlets, and two Higgs doublets respectively. For simplicity, we restrict our attention to one singlet field only in the $i$-summation, which we will denote with $S$.
The covariant derivative acting on the $H_1$ field is defined as
\begin{small}
\ba
{\cal D}_{\mu} H_1 = \left(\partial_{\mu}+i g_2 W^{a}_{\mu} \tau^{a} + \frac{i}{2}g_Y Y_1 A^Y_{\mu} + \frac{i}{2}g_{Z'} B_{H_1} B_{\mu}\right)H_1\nonumber\\
\ea
\end{small}
where $\tau^{a}=\sigma^{a}/2$ and $\sigma^{a}$ are the $SU(2)$ Pauli matrices, $W^{a}_{\mu}$ are the $SU(2)$ gauge fields and $g_2$ is the respective coupling, $A^{Y}_{\mu}$ is the $U(1)_{Y}$ gauge field and $g_Y$ is the coupling, and $B_{\mu}$ is the extra $U(1)_{Z'}$ gauge field and $g_{Z'}$ is the coupling.
$Y_1$ is the hypercharge of $H_1$ under $U(1)_Y$, and $B_{H_1}$ is the charge of $H_1$ under the extra $U(1)_{Z'}$. The $SU(3)$ sector is omitted here.
The covariant derivative acting on the other Higgs fields has the same structure.
We introduce the following parameterisation for the Higgs fields
\ba
&&H_1 = \left(
\begin{array}{c}
\textrm{Re} H_1^0 +i\textrm{Im} H_1^0\\
\textrm{Re} H_1^- + i\textrm{Im} H_1^-
\end{array}
\right),
\nonumber\\
&&H_2  = \left(
\begin{array}{c}
\textrm{Re} H_2^+ +i\textrm{Im} H_2^+ \\
\textrm{Re} H_2^0 +i\textrm{Im} H_2^0
\end{array}
\right),
\nonumber\\
&& S =\textrm{Re} S + i \textrm{Im} S,
\ea
and the VEVs in correspondence of the minimum value of the potential
are defined as
\ba
\langle H_1\rangle = \left(
\begin{array}{c}
v_1 \\
0
\end{array}
\right),
\,\,
\langle H_2\rangle = \left(
\begin{array}{c}
0 \\
v_2
\end{array}
\right),
\,\,
\langle S\rangle=v_S.
\ea
where $v_1$, $v_2$, and $v_S$ are the respective values.
Expanding the scalar sector in Eq.~\ref{scalar-sec} we obtain the quadratic terms that contribute to the gauge bosons mass matrix when the Higgs fields take their VEV's.
These terms are
\begin{eqnarray}
&&{\cal L}_{GM} = \frac{g_{2}^{2}}{4}(v_{1}^{2} + v_{2}^{2})W^{+\mu}W^{-}_{\mu}
+\frac{g_2^{2}}{4}(v_{1}^{2}+v_{2}^{2})W^{3\mu}W^{3}_{\mu}
\nonumber\\
&&\hspace{1cm}
-\frac{g_2 g_{Y}}{4}(v_{1}^{2}+v_{2}^{2})W^{3\mu}A^{Y}_{\mu}
+\frac{g_{Y}^{2}}{4}(v_{1}^{2}+v_{2}^{2})A^{Y\mu}A^{Y}_{\mu}
\nonumber\\
&&\hspace{1cm}
+\frac{g_2 g_{Z'}}{4}(B_{H_{1}}v_{1}^{2}-B_{H_{2}}v_{2}^{2})W^{3}_{\mu}B^{\mu}
\nonumber\\
&&\hspace{1cm}
-\frac{g_{Y}g_{Z'}}{4}(B_{H_{1}}v_{1}^{2}-B_{H_{2}}v_{2}^{2})A^{Y}_{\mu}B^{\mu}
\nonumber\\
&&\hspace{1cm}
+\frac{g_{Z'}^{2}}{4}(B_{H_{1}}^{2}v_{1}^{2}
+B_{H_{2}}^{2}v_{2}^{2} +B_{S}^{2}v_{S}^{2})B_{\mu}B^{\mu},
\end{eqnarray}
where $B_S$ is the charge of the ${S}$ field under the extra $U(1)_{Z'}$. The corresponding mass matrix in the $(W^{3}_{\mu},A^{Y}_{\mu},B_{\mu})$ basis is given by
\ba
{\mathcal M}^2_{gauge}=\left(
\begin{array}{ccc}
\frac{g_2^{2}}{4} v^2 &
-\frac{g_2 g_{Y}}{4}v^2 &
\frac{g_2}{4} x_B\\
\\
-\frac{g_{2} g_{Y}}{4}v^2 &
\frac{g_{Y}^{2}}{4}v^2 &
-\frac{g_{Y}}{4}x_B\\
\\
\frac{g_{2}}{4}x_B &
-\frac{g_{Y}}{4}x_B &
\frac{N_{B}}{4}
\end{array}
\right),
\ea
where we have defined the following quantities to simplify the notation
\ba
&&x_{B} = g_{Z'} (B_{H_1} v_1^2 - B_{H_2} v_2^2)\,,
\nonumber\\
&&N_{B} = g_{Z'}^2 (B_{H_1}^2 v_1^2 + B_{H_2}^2 v_2^2 + B_S^2 v_S^2)\,,
\ea
and where $v^2=v_1^2+v_2^2$, $g^2 = g_Y^2+g_2^2$, and $\tan\beta= v_2/v_1$.
The photon field is massless, and the mass eigenvalues of the $Z$ and the $Z'$ neutral gauge bosons with this notation are
\ba
&&M^2_{Z} =\frac{1}{8}  \left(N_B + g^2 v^2 - \sqrt{ \left(N_B -g^2v^2\right)^2 + 4 g^2 x_B^2}\right)\,,
\nonumber\\
&&M^2_{Z'} =\frac{1}{8}   \left(N_B + g^2 v^2 + \sqrt{ \left(N_B -g^2v^2\right)^2 + 4 g^2 x_B^2}\right)\,.
\nonumber\\
\ea
For large values of $v_S$, the extra $Z'$ boson decouples, and the mass of the $Z$ boson can be expressed as
\ba
M_Z^2=\frac{1}{4} g^2 v^2 - \frac{1}{4}g^2 \frac{\left(B_{H_1} v_1^2 - B_{H_2} v_2^2 \right)^2}{B_S^2 v_S^2}
+ {\cal O}(1/v_S^4),
\nonumber\\
\ea
where $1/4 g^2 v^2$ is the square mass of the SM $Z$ boson, and the other terms are corrections that are $(1/v_S^2)^n$ suppressed.
The mass of the SM $Z$ boson is measured with very high accuracy (i.e. $M_Z = 91.1876\pm 0.0021$ GeV) and it can in principle be used to constrain the value of $v_S$ and $g_{Z'}$ for fixed values of the charges in the model.
\begin{figure}
\begin{center}
\includegraphics[width=7.cm]{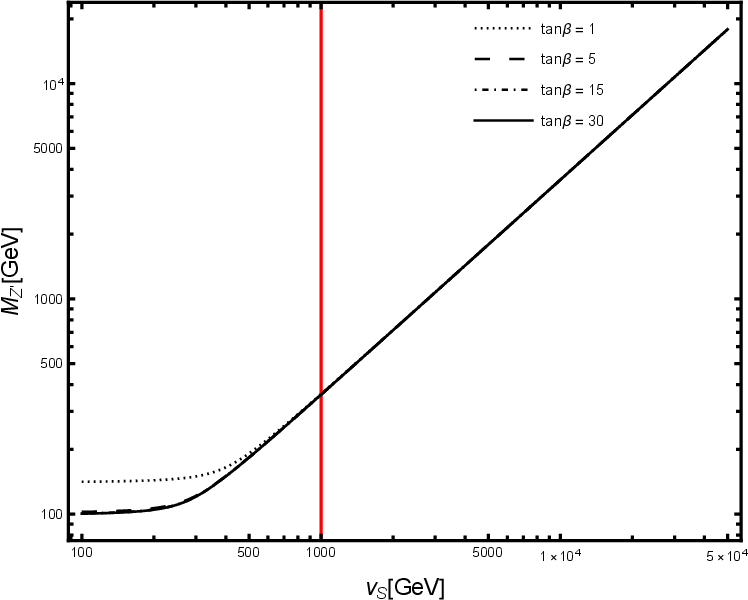}
\caption{TeV-range $Z'$ mass as a function of $v_S$ for different values of $\tan\beta$ and with $g_{Z'}=g_Y$. The region at the right of the red vertical line is the one in which the expectation value $v_S$ is in the TeV range.}
\label{Zp-mass-vs-tanbeta}
\end{center}
\end{figure}

\begin{figure}
\begin{center}
\includegraphics[width=7.cm]{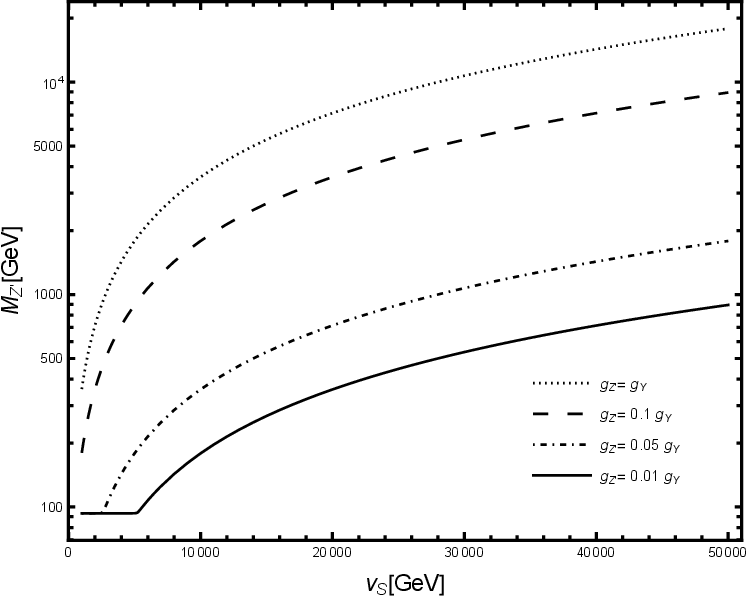}
\caption{TeV-range $Z'$ mass as a function of $v_S$ for different values of the coupling constant $g_{Z'}$. }
\label{Zp-mass-vs-gZp}
\end{center}
\end{figure}
As a simple example, in Figures~\ref{Zp-mass-vs-tanbeta}, and~\ref{Zp-mass-vs-gZp}, we illustrate a particular choice of the parameters $v_S$, $\tan\beta$, and $g_{Z'}$ that leads to values of $M_{Z'}$ in the ${\cal O}(10)$ TeV range. In Figure~\ref{Zp-mass-vs-tanbeta}, $M_{Z'}$ values are obtained as function of $v_S$ for different values of $\tan\beta$ and with $g_{Z'}=g_Y$. The region to the right of the red vertical line is the one in which the expectation value $v_S$ is in the TeV range. We choose $v_S=1$ TeV as a lower limit as $v_S$ is expected to be large. In Figure~\ref{Zp-mass-vs-gZp}, the $v_S$-parameter scan is performed
for different values of the $Z'$ coupling constant $g_{Z'}$. To obtain $M_{Z'}>4$ TeV $v_S$ must be larger than $1.5\times 10^{4}$ GeV.

The normalized orthogonal eigenvector matrix ${\mathcal O}^{gauge}$ rotates the $(W^3_{\mu}, A^{Y}_{\mu},B_{\mu})$ fields of the neutral sector into the physical ones
\ba
\left\{A^{\gamma}_{\mu}, Z_{\mu} ,Z'_{\mu}\right\} = {\mathcal O}^{gauge}.\left\{W^3_{\mu}, A^{Y}_{\mu},B_{\mu}\right\}
\label{Ogauge}
\ea
and has components given in Appendix~\ref{Ogauge-elem}.
The photon field is given by $A^{\gamma}_{\mu}=(g_2 A^{Y}_{\mu} + g_Y W^3_{\mu})/g$. The small mixing between the $Z$ and $Z'$ is expressed through angle $\delta$ defined as
\ba
&&\sin\delta =-\frac{2 g x_B}{\sqrt{8 g^2 x_B^2 + 2f_1^2 (1+\frac{1}{f_1}\sqrt{4g^2 x_B^2+f_1^2}) }}
\nonumber\\
&&\cos\delta = -\frac{f_1 +\sqrt{f_1^2 + 4 g^2 x_B^2}}{\sqrt{8 g^2 x_B^2 + 2f_1^2 (1+\frac{1}{f_1}\sqrt{4g^2 x_B^2+f_1^2})  }}
\nonumber\\
&&\tan{\delta} = \frac{2 g x_B}{f_1 +\sqrt{f_1^2 + 4 g^2 x_B^2}}.
\ea
where we introduced $f_1 = N_B - g^2 v^2$ to simplify the notation.
We observe that when $1/v_S^2$ is small, the structure of the rotation matrix simplifies,
and one has
\ba
&&A^{\gamma}_{\mu}= A^{Y}_{\mu}\cos{\theta_W} +  W^3_{\mu}\sin{\theta_W}
\nonumber\\
&&Z_{\mu} = W^3_{\mu} \cos{\theta_W}- A^{Y}_{\mu}\sin{\theta_W} + B_{\mu}\varepsilon
\nonumber\\
&&Z^\prime_{\mu} = B_{\mu} + \left(\sin{\theta_W} A^{Y}_{\mu} -   W^3_{\mu}\cos{\theta_W} \right)\varepsilon
\ea
where $\theta_W$ is the Weinberg angle, and where $g_2/g = \cos{\theta_W}$, $g_Y/g = \sin{\theta_W}$. The infinitesimal mixing parameter can be defined as
\begin{equation}
\sin\delta|_{v_S\rightarrow\infty} \simeq  -g ~x_B/f_1+{\cal O}(1/v_S^2), ~~~\varepsilon = - g ~x_B/f_1
\label{Ogauge-comp}
\end{equation}
obtained by expanding the expression for $\sin\delta$ in powers of $1/v_S^2$. According to the choice of parameters used in this work, $|\varepsilon|$ approximately varies between $10^{-3}$ and $10^{-5}$ when $0.1\times 10^{4}\leq v_S\leq 2.5\times 10^{5}$ GeV.

\section{The Higgs sector}
\label{Higgs-sector}
The Higgs sector of the superpotential ${\cal W}$ for this model, contains structures of the type
\begin{eqnarray}
{\cal W}&\supset&\lambda\hat{S}\hat{H}_{1}\cdot\hat{H}_{2}+y_{e}\hat{H}_{1}
\cdot\hat{L}\hat{R}+y_{d}\hat{H}_{1}\cdot\hat{Q}\hat{D}_{R}
\nonumber\\
&+&y_{u}\hat{H}_{2}\cdot\hat{Q}\hat{U}_{R},
\end{eqnarray}
where $y_{e}$, $y_{u}$, and $y_{d}$ are parameters and, as mentioned above,
we restrict our attention to the case of one singlet field $\hat S$.
The study of the EW symmetry breaking proceeds similarly to that in Refs.~\cite{Cvetic:1997ky,Coriano:2008xa}.
The scalar potential is given by
\begin{eqnarray}
V&=&\vert\lambda H_{1}\cdot H_{2}\vert ^{2}
+\vert\lambda S\vert ^{2}(\vert H_{1}\vert^{2}+\vert H_{2}\vert^{2})
+\frac{g_2^{2}}{2}\vert H_{1}^{\dagger}H_{2}\vert^{2}
\nonumber\\
&+&\frac{1}{8}(g_2^{2}+g_Y^{2})(H_{1}^{\dagger}H_{1}-H_{2}^{\dagger}H_{2})^{2}
+(a_{\lambda}S H_{1}\cdot H_{2}+h.c.)
\nonumber\\
&+&\frac{g_B^{2}}{8}(B_{H_{1}}H_{1}^{\dagger}H_{1}
+B_{H_{2}}H_{2}^{\dagger}H_{2}+B_{S}S^{\dagger}S)^{2}
+m_{1}^{2}\vert H_{1}\vert^{2}
\nonumber\\
&+&m_{2}^{2}\vert H_{2}\vert^{2}
+m_{S}^{2}\vert S\vert^{2}.
\end{eqnarray}
To have a stable minimum, conditions are obtained by imposing $\partial V/\partial H_i=0$ on the minimum
\ba
&&\left(4 a_\lambda v_2 v_S + B_{H_1} g_{Z'}^2 v_1
\left(B_{H_1} v_1^2 + B_{H_2} v_2^2+B_S v_S^2\right)\right.
\nonumber\\
&&\left.+g^2 v_1 \left(v_1^2-v_2^2\right) + 4 m_1^2 v_1+4 \lambda^2 v_1 v_2^2+4 \lambda^2
 v_1 v_S^2\right) =0
\nonumber\\
\nonumber\\
&&\left(4 a_\lambda v_1 v_S + B_{H_2} g_{Z'}^2 v_2
\left(B_{H_1} v_1^2 + B_{H_2} v_2^2+B_S v_S^2\right)\right.
\nonumber\\
&&\left.+g^2 v_2 \left(v_2^2-v_1^2\right) + 4 m_2^2 v_2+4 \lambda^2 v_1^2 v_2 +4 \lambda^2
 v_2 v_S^2\right) =0
\nonumber\\
\nonumber\\
&&4 a_\lambda v_1 v_2 + v_S \left[4m_S^2 +4 v^2\lambda^2\right]
\nonumber\\
&&+ v_S B_S g_{Z'}^2\left(B_{H_1} v_1^2 + B_{H_2} v_2^2+B_S v_S^2\right)=0
\nonumber\\
\ea
from which one can eliminate $m_1,m_2$ and $m_S$, while $a_\lambda$ and $\lambda$ are free parameters.

\subsection{The CP-even sector}
\label{CP-even-sector}
The matrix elements of the CP-even sector, for which we use the basis $(\textrm{Re}H_1^{0},\textrm{Re}H_2^{0},\textrm{Re}S)$, are obtained from $1/2 \partial^2 V/\partial H_i\partial H_j$ and are given as
\begin{eqnarray}
({\mathcal M}^{2}_{ev})_{11}&=&\frac{1}{2}\left(g_B^{2}B_{H_{1}}^{2}+g^{2}\right)v_{1}^{2}
-a_{\lambda}\frac{v_{2}v_{S}}{ v_{1}}
\nonumber\\
({\mathcal M}^{2}_{ev})_{12}&=&a_{\lambda} v_{S} - \frac{1}{2}v_{1}v_{2}\left(-g_{Z'}^{2}B_{H_{1}}B_{H_{2}} - 4\lambda^{2} + g^{2}\right)
\nonumber\\
({\mathcal M}^{2}_{ev})_{13}&=&a_{\lambda}v_{2}+\frac{1}{2}v_{1}v_{S}
\left(g_{Z'}^2 B_{H_{1}}B_{S}+ 4 \lambda^{2}\right)
\nonumber\\
({\mathcal M}^{2}_{ev})_{22}&=&\frac{1}{2}\left(g_{Z'}^{2}B_{H_{1}}^{2}+g^{2}\right)v_{2}^{2}
-a_{\lambda}\frac{v_{1}v_{S}}{ v_{2}}
\nonumber\\
({\mathcal M}^{2}_{ev})_{23}&=&a_{\lambda} v_{1} + \frac{1}{2}v_{2}v_{S}
\left(g_{Z'}^{2} B_{H_{2}}B_{S}  +4 \lambda^{2}\right)
\nonumber\\
({\mathcal M}^{2}_{ev})_{33}&=&-a_{\lambda}\frac{v_{1}v_{2}}{v_{S}}
+\frac{1}{2}g_{Z'}^{2}B_{S}^{2}v_{S}^{2}.
\end{eqnarray}
This mass matrix is symmetric $({\mathcal M}_{ij}={\mathcal M}_{ji}$ and the other terms are obtained by symmetry.
The diagonalization procedure leads to three massive states corresponding to three neutral Higgs particles indicated by $(H^{0}_{1},H^{0}_{2},H^{0}_{3})$. One of these states is interpreted as the observed Higgs boson with mass $m_H\approx 125$ GeV, one is light ($20 \leq m_H\leq 90$ GeV) and the other has mass in the TeV's range.
The light Higgs states represent a potential new decay channel for the $Z_0$ boson, although this is highly suppressed.

\begin{figure}
\begin{center}
\includegraphics[width=8cm]{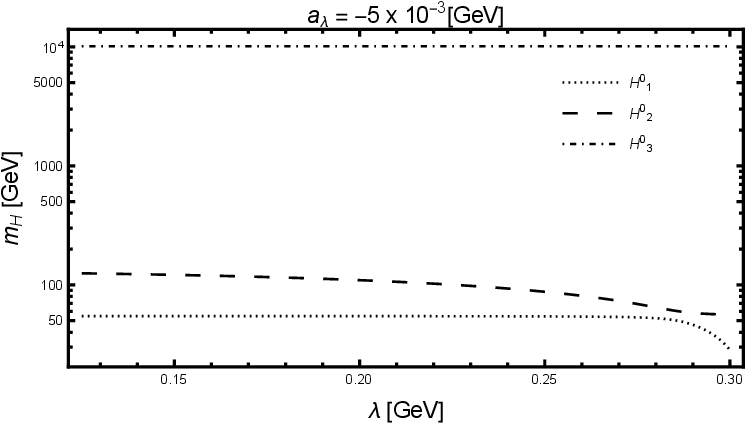}
\includegraphics[width=8cm]{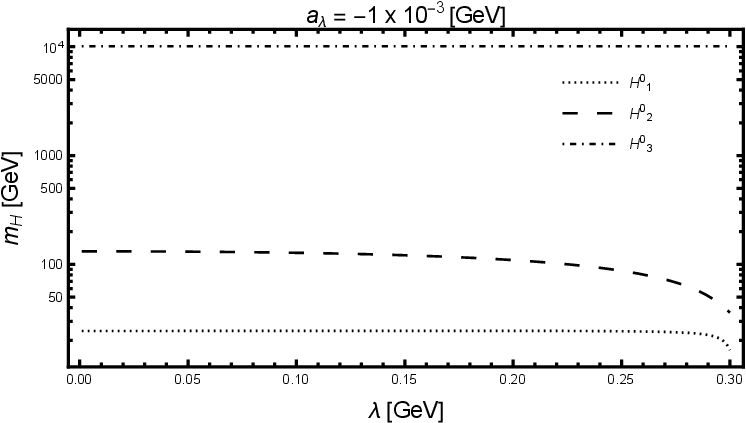}
\caption{Mass values of the CP-even sector Higgs bosons as a function of the scalar potential parameters $\lambda$ for two values of $a_\lambda$. Here we have chosen $\tan\beta=30$, $g_{Z'}=g_Y$, $v_S= 2\cdot 10^{4}$ GeV, and $v_2 =246$ GeV.}
\label{CP-even-higgs-mass}
\end{center}
\end{figure}

As a simple example, the numerical analysis for a selected region of the parameter space for the mass eigenvalues is shown in Figure~\ref{CP-even-higgs-mass}. Here, the Higgs mass values are shown as function of $\lambda$ for two values of $a_\lambda$, and for $\tan\beta=30$, $g_{Z'}=g_Y$, $v_S= 2\cdot 10^{4}$ GeV, and $v_2 =246$ GeV. These values are chosen consistently with those previously discussed in Sec.~\ref{Neutral-sector} based on Figures~\ref{Zp-mass-vs-tanbeta}, and~\ref{Zp-mass-vs-gZp}.
This shows that for relatively small values of the scalar potential parameters $\lambda$ and $a_\lambda$, there exists a portion of the parameter space which leads to acceptable physical mass values for the Higgs states.
The interaction basis states are related to the physical Higgses through the rotation matrix obtained by the inverse matrix of the normalized eigenvectors $\left({\cal U}^{CP-even}\right)^{-1}_{ij}$
\ba
c_{ij} = \left({\cal U}^{CP-even}\right)^{-1}_{ij}
\ea
which results in
\ba
\textrm{Re}H^0_1 &=& c_{11} H^0_1 + c_{12} H^0_2 + c_{13} H^0_3
\nonumber\\
\textrm{Re}H^0_2 &=& c_{21} H^0_1 + c_{22} H^0_2 + c_{23} H^0_3
\nonumber\\
\textrm{Re}S &=& c_{31} H^0_1 + c_{32} H^0_2 + c_{33} H^0_3
\ea
where, in correspondence of the parameters chosen in Figure~\ref{CP-even-higgs-mass}, we obtain $|c_{11}|\approx|c_{22}|\approx|c_{33}|\approx 0.90$, $|c_{12}|=|c_{21}|\approx 10^{-2}$. The remaining coefficients are $\lesssim 10^{-3}$.

\subsection{The CP-odd sector}

For the CP-odd sector we use the $(\textrm{Im} S, \textrm{Im} H^0_1,\textrm{Im} H^0_2)$ basis in correspondence of which
the matrix elements are
\ba
{\mathcal M}^2_{odd}=-a_\lambda\left(
\begin{array}{ccc}
\frac{v_1 v_2}{v_S} & v_2 & v_1\\
v_2 & \frac{v_2 v_S}{v_1} & v_S\\
 v_1 & v_S &\frac{v_1 v_S}{v_2}
\end{array}
\right)
\ea

\begin{figure}
\begin{center}
\includegraphics[width=8cm]{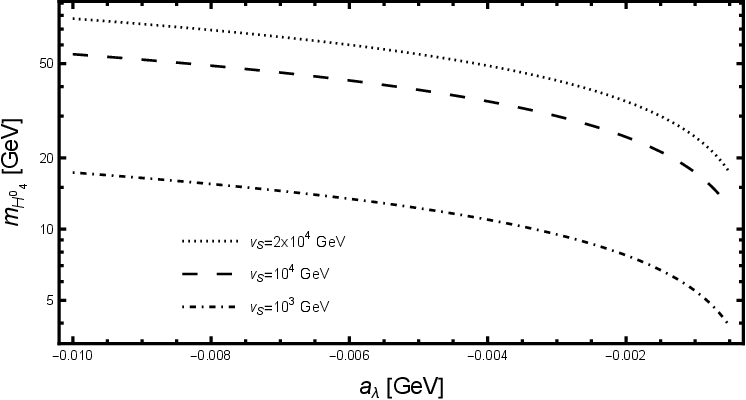}
\caption{Mass values of the CP-odd sector pseudo-scalar Higgs boson
as a function of the scalar potential parameter $a_\lambda$ for different values of $v_S$.
Here we have chosen $\tan\beta=30$ and $v_2 =246$ GeV.}
\label{CP-odd-higgs-mass}
\end{center}
\end{figure}

After the diagonalization procedure we obtain two null eigenvalues, corresponding to two neutral Nambu-Goldstone bosons $G^0_1$ and $G^0_2$, and one physical state $H^0_4$, identified with a neutral pseudoscalar Higgs of mass
\ba
m^2_{H^0_4}= a_\lambda \left(-\frac{v_1 v_2}{v_S} - \frac{v_1 v_S}{v_2} - \frac{v_2 v_S}{v_1} \right)
\ea
Rotating the fields from the interaction basis to the physical basis
in the derivative couplings of the lagrangian density
\ba
{\cal L}_{DC}&=& - g_{2}W^{3}_{\mu}\partial^{\mu}G_Y + g_{Y}A^{Y}_{\mu}\partial^{\mu} G_Y
+ g_{Z'} B_{\mu}\partial^{\mu} G_B
\nonumber\\
\ea
where
\ba
&&G_Y=g_2 (v_{1}\textrm{Im}~H_1^0-v_{2}\textrm{Im}~H_2^0)
\nonumber\\
&&G_B=g_{Z'} (B_{H_{1}}v_{1}\textrm{Im}~H_1^0 + B_{H_{2}}v_{2}
\textrm{Im}~H_2^0+B_{S}v_{S}\textrm{Im}~S),
\nonumber\\
\ea
the two Nambu-Goldstone bosons, $G^0_1$ and $G^0_2$, can be expressed in terms of the physical fields giving $G_Z$ and $G_{Z'}$ which correspond to the $Z$ and $Z'$ massive states.
The expression for ${\cal L}_{DC}$ in terms of physical states is given by
\ba
{\cal L}_{DC}&=& M_{Z} Z_{\mu} \partial^{\mu} G_Z + M_{Z'} Z'_{\mu} \partial^{\mu} G_{Z'}.
\ea
The eigenvectors matrix ${\cal U}^{CP-odd}$ rotates the physical component $H^0_4$ and the $G_Z, G_{Z'}$ fields in the CP-odd sector is given by
\ba
\left(
\begin{array}{c}
H_{4}^{0}\\
G_Z\\
G_{Z'}
\end{array}
\right)=
 {\cal U}^{CP-odd}
\left(
\begin{array}{c}
\textrm{Im}~S\\
\textrm{Im}~H_1^0\\
\textrm{Im}~H_2^0
\end{array}
\right),
\ea
In Figure~\ref{CP-odd-higgs-mass} we show the mass values of pseudo-scalar Higgs as a function of  $a_\lambda$ for different values of $v_S$ and where we have chosen $\tan\beta=30$ and $v_2 =246$ GeV as before. Depending on $v_S$, these mass values approximately range from 10 GeV to 100 GeV.

\subsection{The charged sector}
\begin{figure}
\begin{center}
\includegraphics[width=8cm]{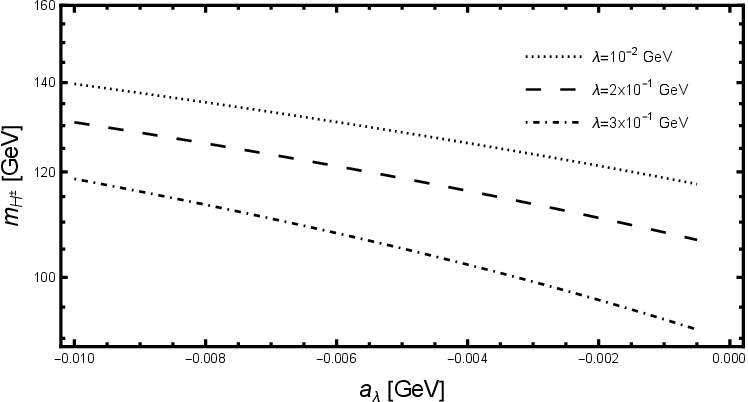}
\caption{Mass values of the charged Higgs as a function of the scalar potential parameter $a_\lambda$ for different values of $\lambda$.
Here we have chosen $\tan\beta=30$, $v_2 =246$ GeV and $v_S=2\cdot 10^{4}$ GeV.}
\label{Chaged-higgs-mass}
\end{center}
\end{figure}

The charged sector is obtained by considering the $(\textrm{Re} H^+_2,\textrm{Re} H^-_1)$
and $(-\textrm{Im} H^+_2,\textrm{Im} H^-_1)$ basis which produces the mass matrix
\ba
{\mathcal M}^2_{H^\pm}=\left(
\begin{array}{ccc}
\frac{v_1 (-2 a_\lambda v_S + v_1 v_2 \kappa)}{2 v_2} &
-a_\lambda v_S + \frac{1}{2} v_1 v_2 \kappa \\
-a_\lambda v_S + \frac{1}{2} v_1 v_2 \kappa &
 \frac{v_2 (-2 a_\lambda v_S + v_1 v_2 \kappa)}{2 v_1}
\end{array}
\right),
\nonumber\\
\ea
where we defined $\kappa = (g_2^2 - 2 \lambda^2)$. Diagonalizing the mass matrix, one obtains a null eigenvalue corresponding to the charged Namubu-Goldstone bosons $G^{\pm}$,
and a mass eigenvalue for the $\pm$ charged states
\ba
m^2_{H^\pm}=(v^2 (-2 a_\lambda v_S + v_1 v_2 \kappa))/(2 v_1 v_2).
\ea
The rotation to the physical basis is obtained by using the eigenvectors matrix ${\cal U}^{charged}$
\ba
\left(
\begin{array}{c}
G^-\\
H^-
\end{array}
\right)=
 {\cal U}^{charged}
\left(
\begin{array}{c}
\textrm{Re} H^+_2\\
\textrm{Re} H^-_1
\end{array}
\right),
\ea
where $G^+=G^{-\dagger}$ and $H^+=H^{-\dagger}$.
In Figure~\ref{Chaged-higgs-mass} we show a plot $m_{H^\pm}$ as a function of $a_\lambda$ for different values of $\lambda$. The other parameter are
chosen as $\tan\beta=30$, $v_2 =246$ GeV, and $v_S=2\cdot 10^{4}$ GeV.

\section{The neutrino sector and sterile neutrinos}
\label{sterile-neutrinos}

\begin{figure}
\begin{center}
\includegraphics[width=8.5cm]{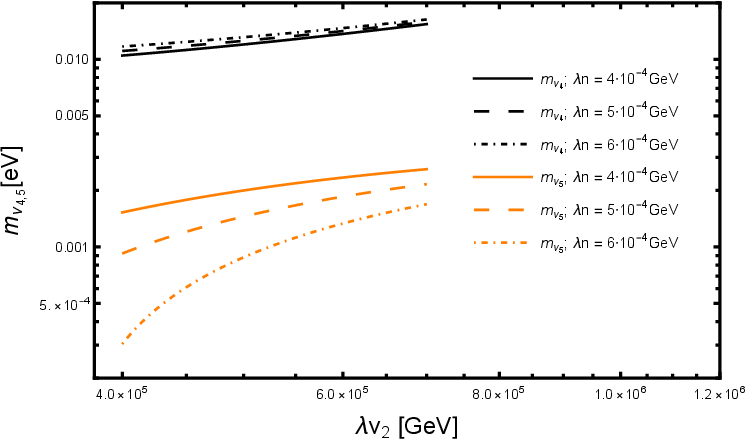}
 \caption{Parametric scan for the mass values of the lightest neutrino states $m_{\nu_{4,5}}$.}
\label{figure1}
\end{center}
\end{figure}

According to the discussion in Sec.~\ref{zprecap} and in~\cite{Faraggi:2018pit}, the seesaw mass matrix that allows for sterile neutrinos with masses of the same size of that of active neutrinos and that have large mixing with the latter, is generated by considering the extended set of fields $\{L^i,N^i,H_{\textrm{vl}}^i,\overline{H}_{\textrm{vl}}^i,\tilde{S}^i,\tilde{H}_1,\tilde{H}_2,\phi,\bar\phi,\zeta^i\}$, which represent the fermionic component of the respective multiplets.
We remind that the model has an underlying $E_6$ structure, and that the fundamental $27$ representation of $E_6$ decomposes as $27={16+\overline{10} +1}$ under $SO(10)$, where
the $16$ contains the SM fields plus the right--handed neutrino field;
the $10$ contains a pair of colour triplets and electroweak doublets; and the $1$ is
a SM and $SO(10)$ singlet field. In total we have at the string level $i=3$, {\it i.e.} three complete multiplets in the $27$ representation of $E_6$, decomposed under
the unbroken $E_6$ subgroup \cite{Faraggi:2014ica}.
Here, $L^i$ are the chiral lepton doublets, $N^i$ are the right-handed neutrinos, $\tilde{S}^i$ are the fermionic components of the $SO(10)$ singlets, and $\tilde{H}_1$ and $\tilde{H}_2$ are the fermionic components of the electroweak Higgses.
The neutrino mass matrix is generated from the allowed renormalizable couplings among these fields obtained by expanding the superpotential in field components
\ba
{\cal W} &\supset&  \lambda_{1}^{ij} \hat{L}^{i} \hat{N}^{j} \hat{H}_2+
\lambda_2^{ijk} \hat{L}^i \hat{N}^j\hat{\bar H}_{\textrm{vl}}^k + \lambda_3^{ij} \hat{N}^i \hat{\zeta}^j\overline{\cal N}
\nonumber\\
&+& \lambda_4^{ijk} \hat{H}_{\textrm{vl}}^i \hat{\bar{H}}_{\textrm{vl}}^j \hat{S}_k
+  \lambda_{5}^{ik} \hat{H}_{\textrm{vl}}^i \hat{H}_2 \hat{S}^{k}+ \lambda_6^{ij} \hat{\bar{H}}_{\textrm{vl}}^i \hat{H}_1  \hat{\zeta}^j
\nonumber\\
&+& \lambda_7^i \hat{H_1} \hat{H}_2 \hat{\zeta}^i
+ \lambda_8^i\hat{\phi} \hat{\bar{\phi}} \hat{\zeta}^i + \lambda_9^i\hat{\bar{\phi}} \hat{\bar{\phi}} \hat{S}^i +\lambda_{10}^{ijk} \hat{\zeta}^i \hat{\zeta}^j \hat{\zeta}^k + \textrm{h.c.}\,,
\label{superpotential}
\nonumber\\
\ea
and from the nonrenormalizable terms $N_i N_j\overline{\cal N}\overline{\cal N}$ introduced to generate a Majorana mass terms for the right–handed neutrinos.
We consider for simplicity only the set of chiral fields under the $U(1)_{Z'}$ group.
Then, the gauge symmetry generates the seesaw mass matrix below
\begin{equation}
\label{seesawmass}
{\cal L}_{M_{\tilde{\nu}}}  =
\begin{pmatrix}
\overline{L}^i \\ \overline{\tilde S}^i\\ \overline{H}_{\textrm{vl}}^i\\ H_{\textrm{vl}}^i\\ \overline{N}^i
\end{pmatrix}^T
\begin{pmatrix}
0 & 0 & 0 & \lambda n & \lambda v_D\\
0& 0 & \lambda v_1 & \lambda v_2 & 0 \\
0  & \lambda v_1  & 0 &v_S &0  \\
\lambda n & \lambda v_2 & v_S & 0 & 0 \\
\lambda v_{D} &0   & 0 & 0 &  \frac{\langle\overline{\cal N}\rangle^2}{M}
\end{pmatrix}
\begin{pmatrix}
L^i \\ \tilde{S}^i\\ H_{\textrm{vl}}^i\\ \overline{H}_{\textrm{vl}}^i\\ N^i
\end{pmatrix}
\end{equation}
which mixes the $(L^i,\tilde{S}^i,H_{\textrm{vl}}^i, \overline{H}_{\textrm{vl}}^i, N^i)$ states.
The VEV $\langle\overline{\cal N}\rangle$ of the heavy Higgs field $\overline{\cal N}$, which breaks the $SU(2)_R$ gauge symmetry 
at a high scale, induces the Majorana mass scale
$\langle\overline{\cal N}\rangle^2/M$ from the  
nonrenormalizable term $N_i N_j\overline{\cal N}\overline{\cal N}$, 
and $M\approx10^{18}$ GeV is related to the heterotic string unification scale. 
The $SU(2)_R$ symmetry can be broken near the GUT or string scale, and is taken here 
as a free parameter. 
The Majorana mass term is large enough for the seesaw mechanism.
The scale $M$ is taken in such a way that the light neutrino masses lie within the current experimental bounds ({\it e.g.}, see refs.~\cite{deSalas:2020pgw,KATRIN:2021uub}) as can be seen from the values shown in Figure~\ref{figure1}. 
To make the phenomenological analysis feasible, we restrict the parameter space by considering
the product of the Yukawa couplings and VEVs as single parameters in Eq.~\ref{seesawmass}. In addition, this allows us to suppress the Yukawa coupling indices appearing in Eq.~\ref{superpotential}.
The seesaw mass matrix in Eq.~\ref{seesawmass}, depends on
$\langle\overline{\cal N}\rangle$ and $n$ that are the VEVs that break
the $SU(2)_R$ symmetry, $v_S$ that is the VEV that breaks the $U(1)_{Z'}$ symmetry,
and $v_{D}$, $v_1$ and $v_2$ which are the VEVs that break the electroweak symmetry.
In particular, $v_{D}$ is the VEV that produces the Dirac mass terms that
couples between the left- and right-handed neutrinos. $v_S$, $v_1$ and $v_2$ have been introduced in Sec.~\ref{Neutral-sector}.

To illustrate the mixing of the states in Eq.~\ref{seesawmass}, we consider the field column
\begin{equation}
n_\alpha^i= \left(L^i,\tilde{S}^i,H_{\textrm{vl}}^i, \overline{H}_{\textrm{vl}}^i, N^i \right)^T.
\label{field-comp}
\end{equation}
We simplify the picture by considering one lepton generation only. To keep our discussion general we retain index $i$, but we fix it as $i=e$. In this way, $\alpha=1,\dots,5$ (the mass matrix is $5\times 5$) and it will be easy to generalize this to the case of three or more generations.

After the diagonalization of the mass matrix in Eq.~\ref{seesawmass}, the mass eigenstates can be written as
\begin{equation}
\tilde{\nu}_{j}  = \left(\nu_{1},\nu_{2},\nu_{3},\nu_{4},\nu_{5}\right)^T  = \sum_{\alpha=1}^5 U^\dagger_{j\alpha} n^e_\alpha,
\end{equation}
where $\nu_{1},\dots,\nu_{5}$ are expressed as combinations of the $L^e,\tilde{S}^e,H_{\textrm{vl}}^e, \overline{H}_{\textrm{vl}}^e, N^e$ fields.
The $\tilde{\nu}$ represents the fermionic component of the sterile neutrino supermultiplet.
In Figure~\ref{figure1} we show the masses of the light neutrino states obtained by the diagonalization procedure as a function of the $\lambda v_2$ parameter for different values of $\lambda n$. These values are compatible with the current experimental findings from global fits of neutrino oscillation data~\cite{deSalas:2020pgw} and mass measurements from the KATRIN collaboration~\cite{KATRIN:2021uub}.

The fermionic Lagrangian in terms of Dirac spinors contains terms of the type
\ba
{\cal L}_{ferm} &\supset& i\bar{L}^i\gamma^{\mu} {\cal D}_{\mu} L^i +  i\bar{R}^i\gamma^{\mu} {\cal D}_{\mu} R^i + i\bar{N}^i\gamma^{\mu} {\cal D}_{\mu} N^i
\nonumber\\
&+& i \bar{\tilde{S}}^i\gamma^{\mu} {\cal D}_{\mu} \tilde{S}^i + \dots
\label{fermion-lagr}
\ea
which is obtained by combining the Weyl spinors of the superfields.
Here, $R^i$ are the right-handed fields relative to $i=e,\mu,\tau$.
It is important to keep in mind that in representation 27 of $E_6$ all chiral multiplets are left-handed (see Table~\ref{table27rot}). The charges for the right-handed fields are obtained from Table~\ref{table27rot} by flipping the respective charge signs.

The neutrino current which couples to the $Z'$ in the physical basis can be written as
\begin{equation}
J^{Z'\overline {\tilde{\nu}}\tilde{\nu}} =  \sum_{i,j} \overline {\tilde{\nu}_{j}} \left[g_{V,ij}\gamma_\mu + g_{A,ij}\gamma_\mu\gamma^5\right]\tilde{\nu}_{i}~Z^{\prime\mu},\\
\end{equation}
where $g_{V,ij}$ and $g_{A,ij}$ are the vector and axial-vector couplings respectively.
They are obtained by rotating $(L^i,\tilde{S}^i,H_{\textrm{vl}}^i, \overline{H}_{\textrm{vl}}^i, N^i)$ through $U^{\dagger}$, and $(W^3_{\mu},A^{Y}_{\mu},B_{\mu})$ through ${\cal O}^{gauge}$ in Eq.~\ref{fermion-lagr}, to the physical states.
In a compact form, $g_{V,ij}$ and $g_{A,ij}$ can be written as
\ba
&&g_{V,ij} = \sum_{\alpha=1}^5 \sum_{k=1}^3 U^{\dagger}_{j\alpha}U_{\alpha i} \left({\cal O}^{T}\right)_{k3} \frac{g_k}{4}(Q_{\alpha,R}+Q_{\alpha,L})_k
\nonumber\\
&&g_{A,ij} = \sum_{\alpha=1}^5 \sum_{k=1}^3 U^{\dagger}_{j\alpha}U_{\alpha i} \left({\cal O}^{T}\right)_{k3} \frac{g_k}{4} (Q_{\alpha,R}-Q_{\alpha,L})_k,
\nonumber\\
\ea
where ${\cal O}^T$ is the transposed of ${\cal O}^{gauge}$ defined in Eq.~\ref{Ogauge}, and in Eq.~\ref{Ogauge-comp} and $(Q_{\alpha,R}\pm Q_{\alpha,L})_k$ are the fermion charges under $SU(2)$, $U(1)_Y$ and $U(1)_{Z'}$ for $k=1,2,3$ respectively, relative to the $\alpha$-th field in Eq.~\ref{field-comp}, and $g_k = {g_2,g_Y,g_{Z'}}$. The $1/4$ coefficient is from the chiral projectors.

% To simplify the notation, we introduce
% \begin{equation}
% \theta_{ij}  = \sum_{\alpha} U^\dagger_{i\alpha} U_{\alpha j}
% \end{equation}
% %

\section{$Z'$ decay rates}
\label{decay-rates}
The $Z'$ decay channels considered in this study are summed up to give the total rate in the equation below
\ba
\Gamma_{Z'} &=& \sum_{f} \Gamma_{Z'\rightarrow f\bar{f}} + \Gamma_{Z'\rightarrow WW}+ \sum_{i=1}^3\Gamma_{Z'\rightarrow Z H^0_i}
\nonumber\\
&+& \sum_{i,j=1}^4\Gamma_{Z'\rightarrow H^0_i H^0_j}+\Gamma_{Z'\rightarrow H^- H^+}+\sum_{i,j=4}^5\Gamma_{Z'\rightarrow \tilde{\nu}_i\overline{\tilde{\nu}_j}}
\nonumber\\
\ea
where $f$ is an index for the quarks and leptons $(e,\mu,\tau)$, $H^0_i$ represents the CP-even ($i=1,2,3$) and CP-odd ($i=4$) states of the two Higgs doublets in the model, $H^\pm$ are the charged Higgs states, and $\tilde{\nu}$ are the mixed neutrino states. Expressions for the individual rates are given in Appendix; here we focus on a $Z'$ decaying into neutrino states as it is more relevant to our discussion about branching ratios.

\begin{figure}
\begin{center}
\includegraphics[width=8cm]{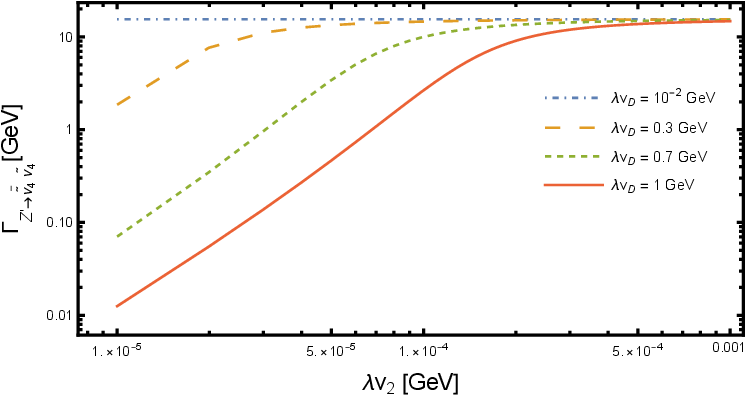}\\
\includegraphics[width=8cm]{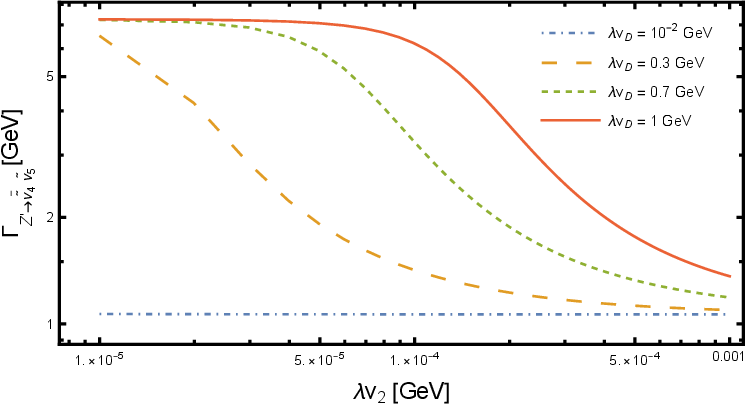}\\
\includegraphics[width=8cm]{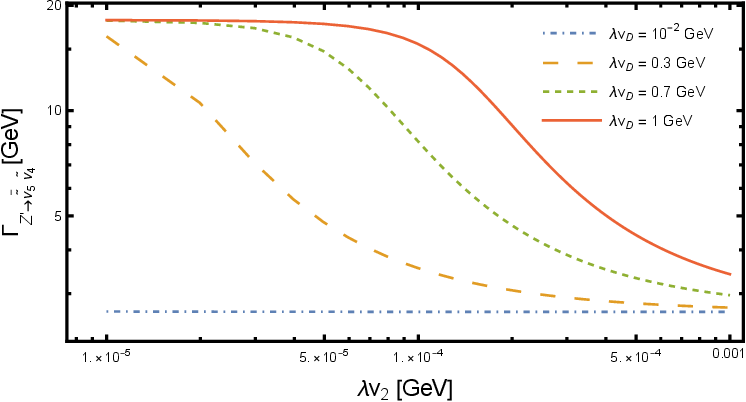}\\
\includegraphics[width=8cm]{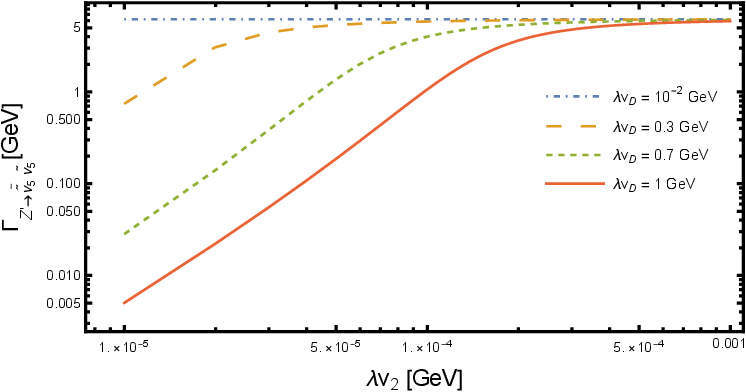}
 \caption{$Z'$ decay rates into $\tilde{\nu}_{i} \overline{\tilde{\nu}}_j$ mixed state with $i,j=4,5$ as a function of $\lambda v_2$ for different values of $\lambda v_D$. $M_{Z'}$ = 7 TeV.}
\label{figure2}
\end{center}
\end{figure}

\begin{figure}
\begin{center}
\includegraphics[width=8cm]{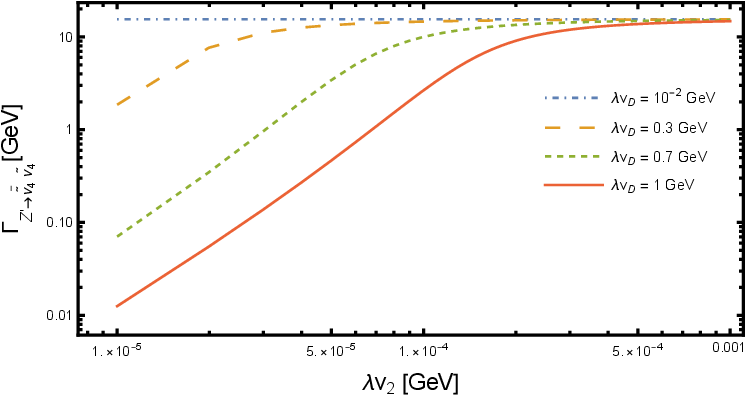}\\
\includegraphics[width=8cm]{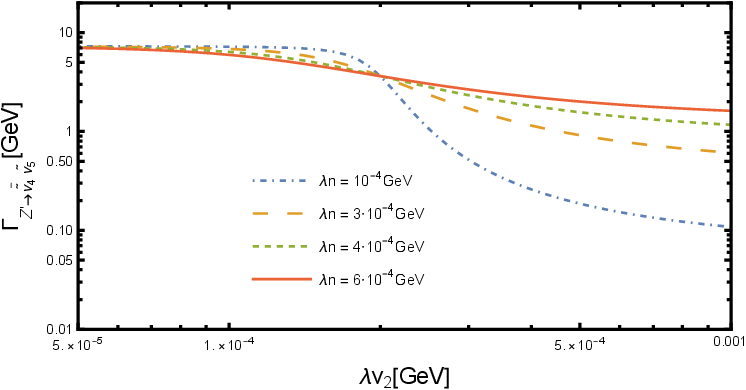}\\
\includegraphics[width=8cm]{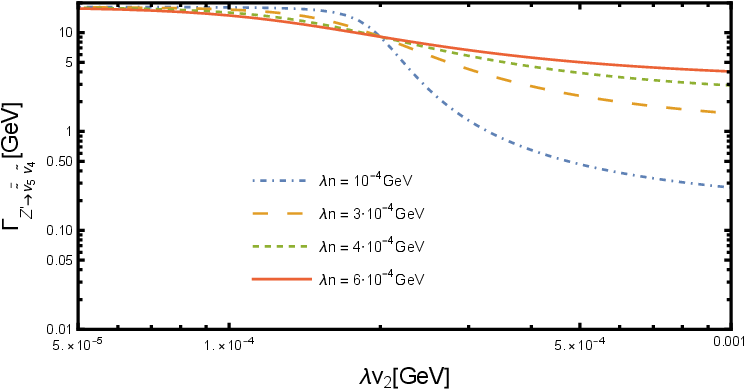}\\
\includegraphics[width=8cm]{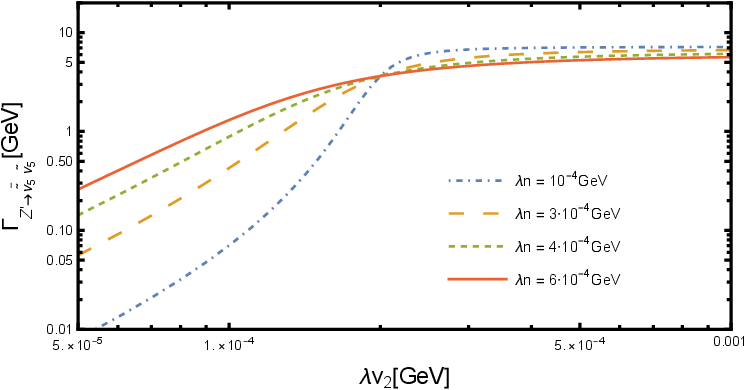}
 \caption{$Z'$ decay rates into $\tilde{\nu}_{i} \overline{\tilde{\nu}}_j$ mixed state with $i,j=4,5$ as a function of $\lambda v_2$ for different values of $\lambda n$. $M_{Z'}$ = 7 TeV.}
\label{figure3}
\end{center}
\end{figure}

\begin{figure}
%\begin{center}
\includegraphics[width=8cm]{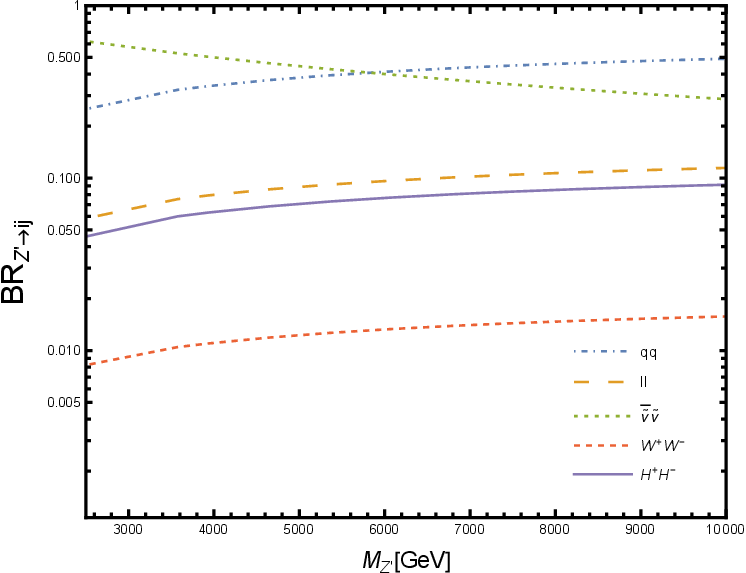}
\caption{$Z'$ branching ratios as a function of $M_{Z'}$ for the most relevant channels: $qq$, $\bar l l$, $\overline{\tilde{\nu}}\tilde{\nu}$, $W^+W^-$, and $H^+H^-$. Here $\lambda v_D = 1$ GeV.}
\label{figure4}
%\end{center}
\end{figure}

The decay rate of the $Z'$ into the allowed neutrino states is given by
\begin{equation}
\Gamma_{Z'\rightarrow i,j} = \frac{M_{Z'}}{12 \pi} \left[g_{V,ij}^2 + g_{A,ij}^2\right],
\end{equation}
which we multiply by a factor of 3 to account for the three generations.
We restrict our attention to the lightest neutrino states $\tilde{\nu}_{4,5}$ that can be considered massless. The remaining states
are very heavy due to the see-saw mechanism, and their decay rate is negligible or zero. In Figures~\ref{figure2} and \ref{figure3} we explored the dependence of $\Gamma_{Z'\rightarrow \overline{\tilde\nu}_{4,5} \tilde\nu_{4,5}}$, on parameters $\lambda n$, $\lambda v_D$, and $\lambda v_2$ which show stronger sensitivity. We keep $v_S=2\cdot 10^4$ GeV which corresponds to a $Z'$ with mass $M_Z'=7$ TeV, and $g_{Z'}=g_Y$. We observe that changes in these parameters produce variations in the decay rates that are from $10^{-3}$ to 15-20 GeV, and therefore considerably impact the total rate of the $Z'$.
The branching ratios for the dominant channels $\bar{q} q$, $\bar l l$, $\overline{\tilde{\nu}}\tilde{\nu}$, $W^+W^-$, and $H^+H^-$, are illustrated in Figure~\ref{figure4} where we observe that the ratio into neutrinos competes with the QCD one $q\bar{q}$.

\section{Collider signatures and $Z'$ mass bounds}
\label{collider-signatures}

In this section we show the results of a phenomenological study for Drell-Yan (DY) dilepton production~\cite{Drell:1970wh} which we used to investigate potential signatures of this model at the LHC. Our notation closely follows that of refs.~\cite{Hamberg:1990np,Coriano:2008wf,Faraggi:2015iaa} and it is briefly summarized below.

The colour-averaged inclusive differential cross section
for resonant dilepton production in proton-proton collisions $p p \rightarrow Z' +X \rightarrow l \bar{l} + X$ at the LHC, is given by~\cite{Hamberg:1990np}
\ba
\frac{d\sigma}{dQ^2}=\tau \sigma_{Z'}(Q^2,M_{Z'}^2)
W_{Z'}(\tau,Q^2)~~~ \tau=\frac{Q^2}{S},
\label{factoriz}
\ea
where $Q$ is the invariant mass of the final-state dilepton pair, $\sigma_{Z'}(Q^2,M_{Z'}^2)$ is the point-like cross section for $Z'$ production (see ref.~\cite{Coriano:2008wf} for details), $\sqrt{S}$ is the center-of-mass energy of the collision, and $W_{Z'}(\tau,Q^2)$ is the hadronic structure function containing details of the hadronic initial-state and the hard-scattering cross section. This is defined as
\ba
W_{Z'}(\tau,Q^2)=\sum_{i,j} \int_{0}^{1}dx_1 \int_0^1 dx_2 \int_{0}^{1}dx
\delta(\tau-x x_1 x_2)
\nonumber\\
f_{H_1\rightarrow i}(x_1,\mu_F^2)f_{H_2\rightarrow j}(x_2,\mu_F^2)\Delta_{i,j}(x,Q^2,\mu_F^2)\,,
\nonumber\\
\ea
where $f_{H_i\rightarrow j}(x_i,\mu_F^2)$ are the parton distribution functions (PDFs) of the proton evaluated at
longitudinal momentum fraction $x_i$ $(i=1,2)$ of the proton ($H_i$) carried by the parton, and at factorization scale $\mu_F$. They represent the probability of finding parton $j$ in hadron $H_i$ at energy scale $\mu_F$ in the collision. The function $\Delta_{i,j}(x,Q^2,\mu_F^2)$ incorporates all the hard-scattering contributions.
This factorization formula in Eq.~\ref{factoriz} is universal for invariant
mass distributions mediated by $s$-channel exchanges of neutral or charged
currents. The hard scatterings can be expanded in a series in terms of the
strong coupling constant $\alpha_s(\mu_R^2)$ as
\ba
\Delta_{i,j}(x,Q^2,\mu_F^2)=\sum_{n=0}^{\infty}
\alpha_s^n(\mu_R^2)\Delta^{(n)}_{i,j}(x,Q^2,\mu_F,\mu_R^2)\,.
\nonumber\\
\ea
where $\mu_R$ is the renormalization scale.
The QCD theory predictions presented here include full spin correlation and are calculated up to next-to-next-to-leading order (NNLO) in $\alpha_s$ with \texttt{CandiaDY}~\cite{Cafarella:2007tj}, an in-house C++ computer code that has been modified to calculate invariant mass distributions for the production of generic $Z'$ models.
\texttt{CandiaDY} has been validated against \texttt{VRAP}~\cite{Anastasiou:2003ds} and \texttt{FEWZ}~\cite{Gavin:2010az,Gavin:2012sy,Li:2012wna}.
Electroweak corrections \cite{Balossini:2006zz,Balossini:2008cs,Baur:2001ze,Zykunov:2005tc,CarloniCalame:2007cd,Arbuzov:2007db,Li:2012wna} are not included in our theory calculation, and this work focuses on invariant mass distribution and total cross section only. More exclusive observables for $Z'$ resonant lepto-production in DY~\cite{Gavin:2012sy,Li:2012wna} will be presented in a forthcoming analysis.

\begin{figure}
%\begin{center}
\includegraphics[width=8.6cm]{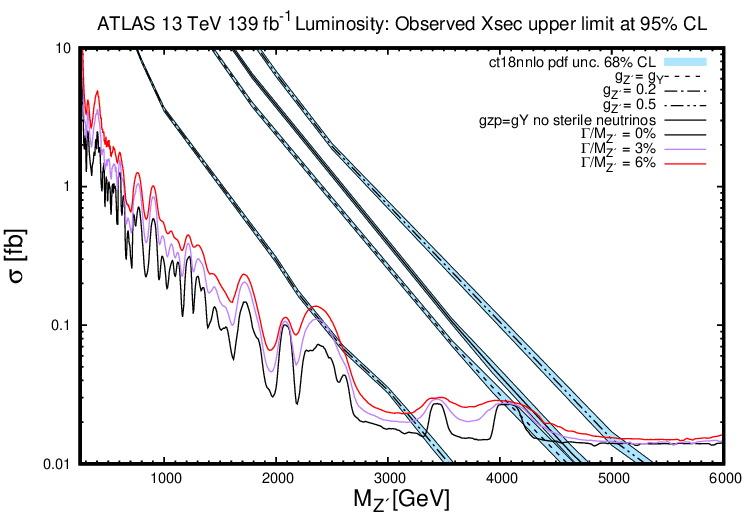}
\caption{Observed 95\% CL upper limits on the total cross section as a function of the $Z'$ mass at ATLAS 13 TeV~\cite{ATLAS:2019erb}. The theory predictions for the total inclusive cross section is evaluated at NNLO in QCD using CT18NNLO PDFs. The induced PDF uncertainty represented by the blue bands is at the 68\% CL.}
\label{figure5}
%\end{center}
\end{figure}

\begin{figure}
%\begin{center}
\includegraphics[width=8.6cm]{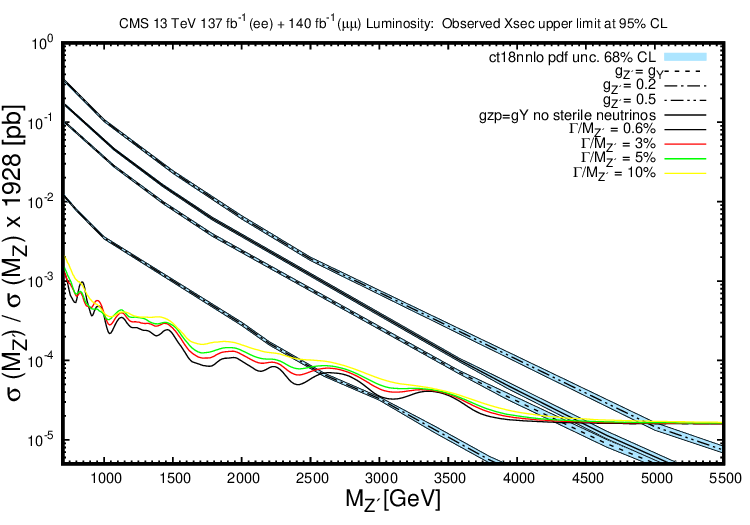}
\caption{Observed 95\% C.L. upper limits on the total cross section ratio
as a function of the $Z'$ mass at CMS 13 TeV~\cite{CMS:2021ctt}.
The theory predictions are evaluated as in Figure~\ref{figure5}.}
\label{figure6}
%\end{center}
\end{figure}

The final results of this phenomenological study are illustrated in Figure~\ref{figure5} and~\ref{figure6} where
we used recent LHC precision measurements at a center of mass energy $\sqrt{S}$ of 13 TeV from the ATLAS~\cite{ATLAS:2019erb} and CMS~\cite{CMS:2018ipm,CMS:2021ctt} experiments for the observed total cross section upper limits at $95\%$ confidence level (CL) as a function of the $Z'$ mass.
On top of these limits, we superimpose our theory calculation at NNLO in QCD calculated with CT18NNLO PDFs~\cite{Hou:2019efy}. We include the induced PDF uncertainties on the total cross sections obtained at 68\% CL which are represented by blue bands in the two figures. Other recent PDF determinations~\cite{Bailey:2020ooq,NNPDF:2021njg} give almost identical results for the cross section calculation.

The observed upper limits at 95\% CL on the production of the same $Z'$ but at CMS 13 TeV~\cite{CMS:2021ctt} are illustrated in~\ref{figure6}.
In this case, to reduce the dependence on correlated systematic uncertainties and also theoretical uncertainties associated to the theory prediction, the CMS limits are expressed in terms of the ratio of the cross section for dilepton production via a $Z'$ boson to the measured cross section for dilepton production via the $Z_0$ boson in the invariant mass range $60–120$ GeV. This ratio is defined as
\ba
\frac{\sigma(M_{Z'})}{\sigma(M_Z)}=\frac{\sigma{\left(pp\rightarrow Z'+ X \rightarrow l\bar{l}+X\right)}}{\sigma{\left(pp\rightarrow Z+ X \rightarrow l\bar{l}+X\right)}}
\ea

The parameter choice utilized to produce our theory predictions for DY $Z'$ production is given below
\ba
&&\{\lambda v_D, \lambda v_1, \lambda v_2, \lambda n, \langle\overline{\cal N}\rangle^2/M\} =
\nonumber\\
&&\{1,5\times 10^{-4},5\times 10^{-4},5\times 10^{-4}, 2.5\times 10^{11}\} ~\textrm{GeV}.
\nonumber\\
&&\tan\beta=30;  ~v_2 = 246 ~\textrm{GeV};~ \mu_F=\mu_R=Q.
\ea
The mass of the $Z'$ is mostly controlled by the vacuum expectation value $v_S$ which varies accordingly for each value of $M_{Z'}$.

The presence of light sterile neutrinos as additional decay channel for the $Z'$ inflates the $Z'$s decay width and reduces the total cross section as compared to the case without sterile neutrinos. This results in a shift of the $Z'$ mass limits towards lighter values. The $Z'$ widths obtained within this model are found to be 1\%-3.5\% of the resonance mass. However, it must be pointed out that they are strongly correlated to the coupling constant $g_{Z'}$ and anti-correlated to the parameter $v_S$, i.e., $\Gamma_{Z'}/M_{Z'}$ increases with $g_{Z'}$ and is suppressed by increasing $v_S$.
As compared to the benchmark models reported in refs.~\cite{ATLAS:2019erb,CMS:2021ctt} which include the
Sequential Standard Model $Z'_{SSM}$~\cite{Altarelli:1989ff} and extra $Z'_{\chi}$ and $Z'_{\psi}$ bosons from $E6$-motivated Grand Unification models \cite{London:1986dk}, the $Z'$ studied in this work has comparable mass limits to that of $Z'_{\psi}$, but systematically lower for $g_{Z'}=g_Y$.

\section{Discussion and conclusions}
\label{conclusions}
In this paper we considered the decay rates of an extra $Z^\prime$ vector boson that arises in the string derived $Z^\prime$ model of ref. \cite{Faraggi:2014ica}. It has been suggested that existence of light
sterile neutrinos at low mass scales can only be reconciled in string
theory provided that there exist an extra $U(1)$ gauge symmetry, under which
the sterile neutrinos are chiral, and which remains unbroken down
to intermediate or low scales \cite{Faraggi:2018pit}.
Taking the $U(1)_{Z^\prime}$
combination from the string model, and assuming unbroken supersymmetry down
to low scales, the spectrum of our string inspired model is nearly fixed
by the requirement of anomaly cancellation. Similarly, its superpotential
is fixed by the symmetries of the model. In this paper, we examined the
decay rate of the extra $Z^\prime$ in the presence of the light sterile
neutrinos and the additional matter states predicted in the model.
It is noted that a priori one may expect that the decay rate of the
extra $Z^\prime$ may be substantially altered in the presence of the
additional light matter states.
We have studied how the presence of sterile neutrinos impacts the
observed cross section upper limits, and consequently, the limits on the mass of the $Z'$. We observed that sterile neutrinos shift the upper limits for the $Z'$ mass towards lighter values.
We have carefully investigated both the neutral gauge bosons and Higgs sectors and studied the dependence of the model on the most important parameters of the superpotential, by restricting the attention to the case of one singlet $S$.
The CP-even sector produces light Higgs states which can couple to the $Z$ and can potentially contribute as new decay channel even if they are highly suppressed.
Finally, we carefully investigated the neutrino sector and explored the rich parameter space of the corresponding sector of the superpotential. These parameters are chosen such that they produce realistic values for the mass of the lighter neutrino states. Sterile neutrinos are the result of a complicated mixture of states as illustrated in Eq.~\ref{seesawmass}, where we restricted the analysis to the case of one lepton generation. Decay rates of the $Z'$ into sterile neutrinos have been explored in terms of the most sensitive parameters $(\lambda n, \lambda v_1,\lambda v_2, v_S, \lambda v_D)$.
To conclude we remark that while our motivation herein was the association
of the sterile neutrinos under which they are chiral, the argument can be extended to the very existence of the electroweak symmetry breaking scale.
From table \ref{table27rot} we note the existence of the electroweak Higgs doublets that
are chiral under $U(1)_{Z^\prime}$ and the additional vector--like pair, which is added in
to facilitate gauge coupling unification \cite{Faraggi:2013nia, Ashfaque:2016jha}.
However, while there is no symmetry that protects the additional vector--like pair
from acquiring a large mass, the mass terms for the chiral Higgs pairs can only
be generated by the $Z^\prime$ breaking VEVs. Thus, we note that the exitence of
the electroweak symmetry breaking scale can be naturally generated in this string inspired
$Z^\prime$ model, with the combination of supersymmetry and the $U(1)_{Z^\prime}$ symmetry. Agreement with the gauge coupling parameters, $\sin\theta_W(M_Z)$ and $\alpha_s{M_Z}$,
then requires a revised scrutiny. We further remark that the existence of the extra
states at the $U(1)_{Z^\prime}$ breaking scale may very naturally give rise to small deviation from the Standard Model predictions, such as the one recently reported by the
CDF--collaboration for the $W$--boson mass \cite{CDF:2022hxs},
without affecting direct searches
as the mass scale of the extra states is naturally associated with the $Z^\prime$ mass scale.
We further note that the model predicts a rich Higgs spectrum and that some of the scalar mass eigenstates may lie below the observed Higgs states at 125 GeV. Recent claims in the literature for evidence for such a state in the LHC data \cite{Biekotter:2022jyr} are therefore particularly intriguing. As a final note, we remark that the $Z^\prime$ model
considered here is supersymmetric. Depending on the details of the supersymmetry
breaking mechanism, the additional superpartner states may, or may not appear, below the
$Z^\prime$ breaking scale. The appealing scenario is in fact the possibility that the supersymmetry breaking scale and the $Z^\prime$ breaking scale are in fact associated.
We will return to these questions in future publications.
\newline
{\bf Acknowledgements.} 
The work of M.G. is supported
by the National Science Foundation under Grant No. PHY-2112025.
AEF is supported in part by a Weston visiting professorship at the Weizmann Institute of 
Science and would like to thank Doron Gepner and the Department of Particle Physics and Astrophysics for hospitality.

\appendix

\section{Rotation Matrix for the neutral gauge boson fields}
\label{Ogauge-elem}
The ${\cal O}^{gauge}$ matrix individual components are reported here
\begin{small}
\ba
&&{\mathcal O}_{11} = g_Y/g,  ~~~~{\mathcal O}_{12} = g_2/g, ~~~~{\mathcal O}_{13} = 0,
\nonumber\\
&&{\mathcal O}_{21} =\frac{-g_2/g ~x_B \left(f_1 + \sqrt{f_1^2 + 4 g^2x_B^2}\right)}{\sqrt{4 g^2 x_B^4 + 2 x_B^2\left(2 g^2 x_B^2+ f_1\left(f_1 + \sqrt{f_1^2 + 4 g^2x_B^2}\right) \right)}} ,
\nonumber\\
&&{\mathcal O}_{22}=\frac{g_Y/g~ x_B \left(f_1 + \sqrt{f_1^2 + 4 g^2x_B^2}\right)}{\sqrt{4 g^2 x_B^4 + 2 x_B^2\left(2 g^2 x_B^2+ f_1\left(f_1 + \sqrt{f_1^2 + 4 g^2x_B^2}\right) \right)}},
\nonumber\\
&&{\mathcal O}_{23}=\frac{1}{\sqrt{2+\frac{ \left(f_1 \left(f_1+\sqrt{f_1^2+4 g^2 x_B^2}\right)\right)}{2 g^2 x_B^2}}} ,
\nonumber\\
&&{\mathcal O}_{31} =\frac{-g_2/g ~x_B \left(f_1-\sqrt{f_1^2 + 4 g^2x_B^2} \right)}{\sqrt{4 g^2 x_B^4 + 2 x_B^2\left(2 g^2 x_B^2+ f_1\left(f_1 - \sqrt{f_1^2 + 4 g^2x_B^2}\right) \right)}} ,
\nonumber\\
&&{\mathcal O}_{32}=\frac{g_Y/g~ x_B \left(f_1 - \sqrt{f_1^2 + 4 g^2x_B^2}\right)}{\sqrt{4 g^2 x_B^4 + 2 x_B^2\left(2 g^2 x_B^2+ f_1\left(f_1 - \sqrt{f_1^2 + 4 g^2x_B^2}\right) \right)}} ,
\nonumber\\
&&{\mathcal O}_{33}=\frac{2}{\sqrt{4+\frac{2 \left(f_1 \left(\sqrt{f_1^2+4 g^2 x_B^2}+f_1\right)+2 g^2 x_B^2\right)}{g^2 x_B^2}}} .
\nonumber\\
\ea
\end{small}
where $x_{B} = g_{Z'} (B_{H_1} v_1^2 - B_{H_2} v_2^2)$, $N_{B} = g_{Z'}^2 (B_{H_1}^2 v_1^2 + B_{H_2}^2 v_2^2 + B_S^2 v_S^2)$, and $f_1 = N_B - g^2 v^2$.

\section{$Z'$ decay rates}

In this section we report the tree-level analytical expressions for the partial decay widths of the $Z'$. In particular, we calculate the rates for $\Gamma_{Z'\rightarrow W^+W^-}$, $\Gamma_{Z'\rightarrow Z H^0_i}$, $\Gamma_{Z'\rightarrow H^0_i H^0_j}$, $\Gamma_{Z'\rightarrow W^\pm H^\mp}$, and  $\Gamma_{Z'\rightarrow H^+ H^-}$.
We refer to \cite{Coriano:2008wf} and references therein for the calculation of the other rates used in this work.

\subsection{$Z'$ decay rates in $W^+W^-$ and $ZH^0_i$}

The calculation for the decay rate $Z'\rightarrow WW$ gives (see also Refs.~\cite{Barger:1987xw,Altarelli:1989ff})
\ba
&&\Gamma_{Z'\rightarrow WW} = \frac{\alpha_{em}}{48}  \cot^2\theta_W   \left[1-4 \left(\frac{M_{W}}{M_{Z'}}\right)^2\right]^{3/2}
\nonumber\\
&&\left(\frac{M_{Z'}}{M_W}\right)^4\left[1 + 20 \left(\frac{M_{W}}{M_{Z'}}\right)^2 + \left(\frac{M_{W}}{M_{Z'}}\right)^4\right]M_{Z'} \varepsilon^2
\nonumber\\
\ea
while that for the $Z'\rightarrow Z H^0_i$ where $H^0_i$ are from the
CP-even sector with $i=1,2,3$ gives
\ba
&&\Gamma_{Z'\rightarrow Z H^0_i} = \frac{g^2_{Z' Z H^0_i}}{16 M_{Z'}^3\pi}
\left(2 + \frac{M_{Z'}^2 +M_Z^2 - m^2_{H_i}}{4 M_{Z'}^2 M_Z^2}\right)\times
\nonumber\\
&& \hspace{1.cm} \sqrt{M_{Z'}^2 -(M_Z+m_{H_i})^2}
\sqrt{M_{Z'}^2 -(M_Z-m_{H_i})^2}
\nonumber\\
\ea
where the couplings are given by
\ba
g_{Z' Z H^0_1} &=& c_{11} v_1 \left[B_{H_1} c_w g_2 g_{Z'}
- c_w^2 g_2^2 \varepsilon + B_{H_1}^2 g_{Z'}^2 \varepsilon
\right.\nonumber\\
&-& \left. g_Y c_w Y_1 \left(B_{H_1} g_{Z'} - 2 c_w g_2 \varepsilon + g_Y s_w \varepsilon Y_1\right) \right]+\dots
\nonumber\\
\nonumber\\
g_{Z' Z H^0_2} &=& c_{22} v_2 \left[-B_{H_2} c_w g_2 g_{Z'}
- c_w^2 g_2^2 \varepsilon + B_{H_2}^2 g_{Z'}^2 \varepsilon
\right.\nonumber\\
&-& \left. g_Y c_w Y_2 \left(B_{H_2} g_{Z'} + 2 c_w g_2 \varepsilon + g_Y s_w \varepsilon Y_2\right) \right] + \dots
\nonumber\\
\nonumber\\
g_{Z' Z H^0_3} &=& c_{33} v_S B_S^2  g_{Z'}^2  \varepsilon +\dots\,
\ea
where the dots indicate terms proportional to off-diagonal elements $c_{ij}$ that are suppressed and essentially have no impact on the rates (see Sec.\ref{CP-even-sector} for discussion). Here we define
$c_w = \cos{\theta_W}$ and $s_w = \sin{\theta_W}$ to simplify the notation.

\subsection{$Z'$ decay rates in two Higgs bosons}

The decay rate for the $Z'\rightarrow H^0_i H^0_j$ with $i,j=1,2,3,4$ is given by
\ba
&&\Gamma_{Z'\rightarrow H^0_i H^0_j} = \frac{g_{Z'H^0_i H^0_j}^2}{(16 M_{Z'}^5 \pi)}
\left[(m_{H_i} - m_{H_j} - M_{Z'})\times\right.
\nonumber\\
&&\hspace{1.5cm}\left. (m_{H_i} + m_{H_j} - M_{Z'})(m_{H_i} - m_{H_j} + M_{Z'})\right.
\nonumber\\
&&\left.
\hspace{1.5cm} (m_{H_i} + m_{H_j} + M_{Z'})\right]^{3/2};
\ea
where the couplings are defined as
\ba
&&g_{Z' H^0_1 H^0_i} = B_S ~c_{31} ~c_{3i} ~ g_{Z'}~~~i=1,2,3;
\nonumber\\
&&g_{Z' H^0_2 H^0_j} = B_S~ c_{32}~ c_{3j} ~ g_{Z'}~~~j=2,3;
\nonumber\\
&&g_{Z' H^0_3 H^0_3} = B_S~ c_{33}^2 ~ g_{Z'};
\nonumber\\
&&g_{Z' H^0_4 H^0_i} = \frac{1}{2} b_{11}~ B_S ~c_{3i} ~g_{Z'}~~~i=1,2,3;
\ea
where $c_{ij}$ are obtained numerically. They are the coefficients of the inverse matrix of the normalized eigenvectors for the CP-even sector and are discussed in Sec.\ref{CP-even-sector}. To simplify the notation, we introduced the $b_{11}$ coefficient which is defined as
\ba
b_{11} = \frac{v_1^2 v_S \sqrt{v_2^2 + v_S^2}}{v_2^2 v_S^2  + v_1^2 (v_2^2 + v_S^2)}
\ea
\newline
\subsection{$Z'$ decay rates into $W^\pm H^\mp$}

The decay rate of the $Z'$ into $W^+ H^-$ is given by
\ba
&&\Gamma_{Z'\rightarrow W^+ H^-} = \frac{g^2_{Z' W^+ H^-}}{16 M_{Z'}^3 \pi}
\left[2 + \frac{(M_{Z'}^2+ M_W^2-m^2_{H^\pm})}{4 M^2_{Z'} M_W^2}\right] \times
\nonumber\\
&&\sqrt{\left(M^2_{Z'} - (M_W+m^2_{H^\pm})^2 \right)
\left(M^2_{Z'} -(M_W - m^2_{H^\pm})^2\right) }
\nonumber\\
\ea
The rate into $W^- H^+$ is the same. The $g^2_{Z' W^\mp H^\pm}$ couplings are defined as
\ba
g_{Z' W^\mp H^\pm} &=& \frac{1}{\sqrt{2}} g_2 \left(-\frac{v_2}{v} B_{H_1} g_{Z'} v_1 + \frac{v_1}{v} B_{H_2} g_{Z'} v_2 \right.
\nonumber\\
&-&\left. \frac{v_2}{v} g_Y s_w v_1\varepsilon Y_1 + \frac{v_1}{v} g_Y s_w v_2\varepsilon Y_2\right).
\ea
 
\subsection{$Z'$ decay rates into $H^+ H^-$}
The decay rate of the $Z'$ into $H^+ H^-$ is obtained as
\ba
&&\Gamma_{Z'\rightarrow H^+ H^-} = \frac{g^2_{Z' H^+ H^-}}{16 M^5_{Z'} \pi^2}
\left[M^2_{Z'} (M_{Z'} - 2 m_{H^\pm})\right.
\nonumber\\
&&\hspace{2cm}\left.(2 m_{H^\pm} + M_{Z'})\right]^{3/2}\,
\ea
where the coupling is given by
\ba
g_{Z' H^+ H^-} &=& \left(\frac{v_1}{v}\right)^2 \left(B_{H_1} g_{Z'} + c_w g_2\varepsilon\right)
\nonumber\\
&+&\left(\frac{v_2}{v}\right)^2 \left(-B_{H_2} g_{Z'} + c_w g_2\varepsilon\right)
\nonumber\\
&+&g_Y s_w \varepsilon \left(\left(\frac{v_1}{v}\right)^2 Y_1 - \left(\frac{v_2}{v}\right)^2 Y_2\right);
\nonumber\\
\ea

%\bibliography{paperbib}
\bibliographystyle{apsrev4-1}

\end{document}